\documentclass[aps,preprint]{revtex4-1}
\usepackage{graphicx}
\begin{document}

\title{Separability and hidden symmetries of Kerr-Taub-NUT spacetime  in Kaluza-Klein theory}

\author{\large G\"{o}ksel Daylan  Esmer}
\address{Istanbul University,  Department of Physics,
Vezneciler, 34134 Istanbul, Turkey}

\date{\today}

\begin{abstract}

The Kerr-Taub-NUT spacetime in the Kaluza-Klein theory represents a  localized stationary and axisymmetric  object in four dimensions from the Kaluza-Klein  viewpoint. That is, it harbors  companion electromagnetic and  dilaton fields, thereby showing up the signature of the extra fifth dimension. We explore the separability structure of this spacetime and show that the Hamilton-Jacobi equation for geodesics admits the complete separation of variables only for {\it massless} geodesics. This implies the existence of the hidden symmetries in the spacetime, which are generated by the {\it conformal} Killing tensor. Using a simple trick  built up on a conformally related metric (an ``effective" metric) with the Killing tensor, we construct the explicit expression for the conformal Killing tensor.

\end{abstract}

\maketitle

\newpage

\section{Introduction}

It is long known that the Einstein vacuum field equations  admit an intriguing exact solution that describes  a localized stationary and axisymmetric object, the so-called rotating  {\it gravitational dyon}. In other words, general relativity (GR) admits a simple generalization of the ordinary  Kerr  spacetime to include the ``magnetic mass", in addition to the ordinary mass \cite{demianski} (see also  Refs.\cite{newman, misner}). This spacetime is known as the Kerr-Taub-NUT (Newman-Unti-Tamburino) solution of GR. The presence of the magnetic  mass or the NUT charge in the spacetime   results in  conical singularities on its axis of symmetry, thereby making the spacetime asymptotically nonflat. This in turn  enables one to elaborate on
the gravitomagnetic analogue of Dirac's string quantization condition \cite{misner}. Thus, one can think of the NUT charge as  ``dual" to the ordinary mass in GR  just as the electric charge can be thought of as dual to the magnetic charge in the theory of electromagnetism. Continuing this line of the interpretation, one  can also consider the conical singularities as the physical source of a  string which lies  on the axis of symmetry  of the black hole spacetime \cite{bonnor}. On the other hand,  the conical singularities can be removed by imposing a periodicity condition on the time coordinate, but it creates another pathology in the spacetime, namely, the appearance of closed timelike curves (CTCs). Thus, in general, it  is  hard to interpret the Kerr-Taub-NUT  spacetime  as  describing a regular rotating  black hole, in contrast to the  original Kerr  spacetime.

In addition to these undesired features, the Kerr-Taub-NUT  spacetime also possesses a number of remarkable properties.  Among them, the separability property  is of particular interest \cite{carter, bini1, bini2}. It turns out this spacetime shares  the hidden symmetries of the ordinary Kerr spacetime, generated by  a second-rank Killing  tensor \cite{wp}, thereby  rendering both the Hamilton-Jacobi and Klein-Gordon equations  completely integrable. Due to these properties, the Kerr-Taub-NUT spacetime has been an attractive model for exploring various gravitomagnetic  phenomena in asymptotically nonflat spacetimes (see \cite{bini1,  bini2, radu, aliev1} and references therein). It is important to note  that the spacetimes with the NUT charge  turn out to be  an inseparable part of the low-energy string theory as well. Here one can show that the duality symmetries of the effective action allow to construct new stationary Taub-NUT type solutions to the theory \cite{jp,gk}. The NUT charge also provides  a kind of new electric-magnetic duality that was revealed in supersymmetric configurations of the Kerr-Newman-Taub-NUT-AdS spacetime \cite{ortin}. The spacetimes  with NUT charges have also been studied in higher-dimensional GR  with a cosmological constant, where the most general Kerr-AdS  solution with multi-NUT charges  was found in \cite{chen1}.  It was shown that this spacetime admits  a closed conformal  Killing-Yano $2$-form which encodes all its hidden symmetries \cite{fk1, fkrev}.

The investigation of the Kerr-Taub-NUT spacetime is also of interest from the Kaluza-Klein viewpoint. In \cite{aht}, it was shown that the Kerr-Taub-NUT solution in the  Kaluza-Klein theory carries the signature of the extra fifth dimension,  acquiring the electric and dilaton charges though the dilaton  charge is not an independent parameter. This solution was obtained by employing a well-known solution generating procedure \cite{gw, fz} that in essence involves  uplifting the original Kerr-Taub-NUT  metric to five dimensions and  boosting it  in the flat fifth dimension with a subsequent Kaluza-Klein reduction  to four dimensions. Clearly, the resulting metric  satisfies the coupled Einstein-Maxwell dilaton field equations in four dimensions.

Meanwhile, such a procedure of the construction of the solution raises a natural question: {\it Are the separability property and the associated hidden symmetries of the original  spacetime  preserved  by the solution generating procedure}? In a recent work \cite{ag},  this question  was explored for the Kerr-Kaluza-Klein black hole. In this Letter, we continue this line of the exploration and  address the same question for Kerr-Taub-NUT spacetime  in  the Kaluza-Klein theory. In Section 2 we briefly  discuss the spacetime metric and show that the Hamilton-Jacobi equation for geodesics in the spacetime  under consideration exhibits  the  complete separability structure  only for massless  geodesics. Clearly, it is the conformal Killing tensor that generates the hidden symmetries underlying the separability in the massless case. In Section 3 we employ a simple trick, which involves a conformally related metric with the pertaining Killing tensor, and we give the explicit expression for the conformal Killing tensor.

\section{the Metric and the Hamilton-Jacobi Equation}

We recall that the Kerr-Taub-NUT spacetime  in the Kaluza-Klein theory  represents a localized  rotating object with a NUT charge in four dimensions from the Kaluza-Klein  viewpoint and involves the Maxwell and dilaton fields.
The details of the construction of the spacetime metric can be found in \cite{aht}. Here we will only  focus on some basic steps of the construction. The first step amounts to uplifting the original Kerr-Taub-NUT   solution \cite{demianski} to five dimensions, by adding to it a flat spacelike dimension. This results in the five-dimensional metric which in  the Boyer-Lindquist coordinates is given by
\begin{eqnarray}
ds_{5}^2 & = & -{{\Delta}\over {\Sigma}} \left(dt - \chi\,
d\phi \right)^2 + \Sigma \left(\frac{dr^2}{\Delta} +
d\theta^{\,2}\right)
+ \,\frac{\sin^2\theta}{\Sigma} \left[a  dt -
\left(r^2+a^2+\ell^2\right) d\phi\, \right]^2 + dy^2,
\label{5ktnut}
\end{eqnarray}
where the metric functions
\begin{eqnarray}
\Delta &= & r^2 - 2M r + a^2 - \ell^2 \,\,, ~~~~~
\Sigma =  r^2 + (\ell + a \cos\theta)^2 \,\,,\nonumber \\[2mm]
\chi & = & a \sin^2\theta - 2 \ell \cos\theta \,,
\label{metfunct}
\end{eqnarray}
and  $ M $  is the mass parameter, $ a \,$  is the rotation parameter,  $ \ell \, $ is the NUT charge. At the second step, one needs to boost this metric in the fifth dimension, applying the transformation
\begin{eqnarray}
t & \rightarrow &  t \cosh\alpha + y \sinh\alpha \nonumber \\[2mm]
y & \rightarrow &  y \cosh\alpha + t \sinh\alpha\,
\label{boost}
\end{eqnarray}
with the boost velocity  $ v= \tanh\alpha $. Putting now the resulting    metric  into the standard  Kaluza-Klein form
\begin{eqnarray}
ds_{5}^2 &=&   e^{-2 \Phi/\sqrt{3}}\,  ds_{4}^2 + e^{4 \Phi/\sqrt{3}}\,\left(dy + 2 A\right)^2,
\label{kkmetric}
\end{eqnarray}
and performing a compactification along the fifth dimension, one obtains  the four-dimensional metric
\begin{eqnarray}
ds_{4}^2 & = & -\frac{1}{B}\,{{\Delta}\over {\Sigma}} \left(\,dt - \chi \cosh\alpha \,d\phi\,\right)^2 + B \Sigma \left(\frac{dr^2}{\Delta}\,+\,
d\theta^{\,2}\right) -\frac{\Delta \sin^2\theta}{B} \sinh^2\alpha\, d\phi^2
\nonumber\\[2mm] &&
+ \,\frac{\sin^2\theta}{B \Sigma} \left[a dt -
\left(r^2+a^2+\ell^2\right)\cosh\alpha\, d\phi\, \right]^2,
 \label{solution2}
\end{eqnarray}
which is accompanied by the  potential one-form $ A $  and by the dilaton field $\Phi$. We have
\begin{eqnarray}
A &=& \frac{Z \sinh\alpha}{2 B^2}\,\left[\cosh\alpha \, dt - \left( \chi + \frac{ 2 \ell \cos\theta }{Z}\right)d\phi\right],~~~~ \Phi=\frac{\sqrt{3}}{2} \ln B\,,
\label{potform1}
\end{eqnarray}
where
\begin{eqnarray}
B &=& \left(1+ Z \sinh^2\alpha\right)^{1/2}\,,~~~~~~~ Z =   2 \,\frac{ M r + \ell \left(\ell+ a \cos\theta \right)}{\Sigma}\,.
\label{b}
\end{eqnarray}
It is straightforward to check that  this solution  satisfies the equation of motion derived from the four-dimensional action of the  Kaluza-Klein theory
\begin{eqnarray}
S &=&  \int d^4 x \sqrt{-g} \left[R - 2 \left(\partial\Phi\right)^2 - e^{2 \sqrt{3}\, \Phi} F ^2\right],
\label{4d action}
\end{eqnarray}
where  $ F=dA $.  Meanwhile, the physical parameters  of the metric; the total mass, angular momentum  and the total electric charge are given by
\begin{eqnarray}
\mathcal{M} &=& \frac{M}{2} \left(\frac{2- v^2}{1-v^2}\right),~~~~~J = \frac{a M}{\sqrt{1-v^2}} \,\,, ~~~~~Q = \frac{M v }{1-v^2}\,.
\label{mjq1}
\end{eqnarray}
As for the dilaton charge, it is not independent as one can express it in terms of the other parameters (see Refs. \cite{gw, aht}  for details). Thus, the spacetime metric in (\ref{solution2}) generalizes the  Kerr-Taub-NUT  solution of GR to include the effects of the  extra dimension that show up through the appearance of the electric and dilaton charges in the spacetime. For $\ell=0 $,  this solution  goes over into the  boosted Kerr black hole metric  that was earlier  found in  \cite{fz}.

Let us now consider the Hamilton-Jacobi equation  for geodesics in the background of  spacetime   (\ref{solution2}). It is given by
\begin{equation}
\frac{\partial S}{\partial \lambda}+\frac{1}{2}\,g^{\mu\nu}\frac{\partial S}{\partial x^\mu}\frac{\partial S}{\partial x^\nu}=0 \,,
 \label{HJeq}
\end{equation}
where $ \lambda  $ is  an affine parameter.  Using the fact the spacetime under consideration admits two commuting timelike and spacelike Killing vectors one can assume that
\begin{equation}
S=\frac{1}{2}\,m^2\lambda - E t+L \phi +F(r,\theta)\,,
\label{ss}
\end{equation}
where $ F(r,\theta) $ is an arbitrary function. The constants of motion  are the  mass  $ m $,   the total energy $  E $ and  the angular momentum  $ L $  of the particle. Substituting this action  together with the  contravariant components of metric  (\ref{solution2}), given by
\begin{eqnarray}
g^{00}&=&\frac{1}{B\Sigma}\left[\Sigma \sinh^2\alpha+ \frac{\chi^2 \cosh^2\alpha}{\sin^2\theta}-\frac{(r^2 + a^2 +\ell^2)^2\cosh^2\alpha}{\Delta}\right],\nonumber
\\[3mm]
g^{11}&=&\frac{\Delta}{B\Sigma}\,,~~~~~~g^{22}=\frac{1}{B\Sigma}\,,~~~~g^{03}= \frac{\cosh\alpha}{B\Sigma}\left[\frac{\chi}{\sin^2\theta}- \frac{a}{\Delta}\,(r^2 + a^2 +\ell^2)\right], \nonumber \\[3mm]
g^{33}&=& \frac{1}{B\Sigma}\left(\frac{1}{\sin^2\theta}- \frac{a^2}{\Delta}\right)
\label{contras}
\end{eqnarray}
into equation (\ref{HJeq}),  we reduce it into the form
\begin{eqnarray}
&& \Delta  \left(\frac{\partial F}{\partial r}\right)^2 + \left(\frac{\partial F}{\partial \theta} \right)^2 +\left[\Sigma \sinh^2 \alpha + \left(\frac{\chi^2}{\sin^2\theta}- \frac{(r^2 + a^2 +\ell^2)^2}{\Delta}\right)\cosh^2\alpha \right]E^2
\nonumber
\\[2mm] &&
+\left(\frac{1}{\sin^2\theta}- \frac{a^2}{\Delta}\right)L^2 -2\cosh\alpha\left[\frac{\chi}{\sin^2\theta}- \frac{a}{\Delta}\,(r^2 + a^2 +\ell^2)\right]E L = -m^2 B \Sigma\,.
\label{eq1}
\end{eqnarray}
With equation (\ref{b}) in mind, it is not difficult to see that separation of the $ r $  and  $\theta $ variables in this equation does not occur due to the presence of the function $ B $ on the right-hand side. Meanwhile, such a separation  does occur for massless geodesics ($ m=0 $). As is known,  this fact of separability  guarantees the existence of  a new  conserved quantity along the  null geodesics. It is the second-rank symmetric  conformal Killing tensor that generates the  hidden symmetries underlying the separability in the massless case \cite{wp}. In what follows, we  explore the hidden symmetries and  construct the explicit  form  for  the conformal Killing tensor.

\section{The Conformal Killing Tensor}

In a recent work \cite{ag}, the conformal Killing tensor and  the associated
hidden symmetries of  the Kerr-Kaluza-Klein  black hole were explored
by using a simple trick  based on a conformally related metric with the Killing tensor. Here we  shall use similar trick  for constructing the conformal Killing tensor of the the Kerr-Taub-NUT spacetime  in the Kaluza-Klein theory. As it was shown above, the associated spacetime metric  does not allow the full separation of variables for the massive Hamilton-Jacobi equation. That is, the spacetime does not admit  the Killing tensor $ K_{\mu\nu} $ given by the equation
\begin{equation}
\nabla_{(\lambda} K_{\mu\nu)}=0\,,
\label{Kteq}
\end{equation}
where $\nabla $  denotes the covariant differentiation  with respect to the spacetime metric $ g_{\mu\nu} $. On the other hand, one can assume that such a Killing tensor  exists for a conformally related metric $ h_{\mu\nu} $ (an effective  metric) given as
\begin{eqnarray}
h_{\mu \nu} &= & e^{2\Omega} g_{\mu \nu}\,,
\label{confmet}
\end{eqnarray}
where $ \Omega $ is a scalar function and $ g_{\mu\nu} $ is supposed to be the original metric  in (\ref{solution2}). It is not difficult to show that
the associated covariant derivatives of a second-rank symmetric tensor $ P_{\mu\nu} $ are related by
\begin{equation}
D_{(\lambda} P_{\mu\nu)} =\nabla_{(\lambda} P_{\mu\nu)} - 4 P_{(\mu\nu}\Omega_{,\,\lambda)}- g_{(\mu\nu} I_\lambda)\,,
\label{condeq}
\end{equation}
where
\begin{equation}
 I_\lambda= -2\, g^{\alpha\tau} P_{\lambda \tau}\,\Omega_{,\,\alpha}\,.
\label{cur}
\end{equation}
The operator $ D $ denotes covariant differentiation with respect to the metric $ h_{\mu\nu} $  and the comma  stands for the partial derivative. Next, it is straightforward to show that with the tensor $  P_{\mu\nu} $  given in the form
\begin{eqnarray}
P_{\mu \nu} &= & e^{-4\Omega} K_{\mu \nu}\,,
\label{conftens}
\end{eqnarray}
where $  K_{\mu\nu} $ is the Killing tensor for the effective metric   $  h_{\mu\nu} $, equation (\ref{condeq}) can be cast to the defining equation for the conformal Killing tensor. That is, we have
\begin{equation}
\nabla_{(\lambda} P_{\mu\nu)} = g_{(\mu\nu} I_\lambda)\,.
\label{confkil}
\end{equation}
On the other hand, taking the scalar function $ \Omega $  as
\begin{equation}
\Omega = -\frac{1}{2}\ln B\,,
\label{omega}
\end{equation}
one can  show that  the massive Hamilton-Jacobi equation in the background of the effective metric $  h_{\mu\nu} $ admits the complete separation of variables. In this case, the factor $ B $  on the right-hand side of equation (\ref{eq1}) disappears  and for the action $ S $  in the form
\begin{equation}
S=\frac{1}{2}m^2\lambda - E t+L \phi +S_r(r)+S_\theta(\theta)\,,
\label{ss1}
\end{equation}
we obtain two independent ordinary differential equations given by
\begin{eqnarray}
\label{eqr1}
&& \Delta \left(\frac{dS_r}{d r}\right)^2 - \frac{1}{\Delta}\left[(r^2 + a^2 +\ell^2)\cosh\alpha \,E -a L\right]^2 + r^2 \left(m^2 +\sinh^2\alpha\, E^2 \right) = -K\,,
\\[4mm] &&
\left(\frac{dS_\theta}{d \theta}\right)^2 +\frac{1}{\sin^2\theta}\left(\chi \cosh\theta\, E- L\right)^2 + (\ell + a \cos\theta)^2 \left(m^2+ \sinh^2\alpha\, E^2\right)= K\,,
\label{eqth1}
\end{eqnarray}
where $ K $ is a constant of separation.  It is clear that  the separability implies the existence of a new quadratic integral of motion $ K=K^{\mu\nu }p_{\mu} p_{\nu} \,$. This is associated with the hidden symmetries of the effective metric  $  h_{\mu\nu} $, which are  generated by the irreducible  Killing tensor  $  K^{\mu\nu} $. Taking this into account in equation (\ref{eqth1}), with the normalization condition $  m^2= -h^{\mu\nu }p_{\mu} p_{\nu} \,$ in mind, we find the explicit form for the Killing tensor. We have
\begin{eqnarray}
&&K^{\mu\nu}=  \delta^\mu_\theta \delta^\nu_\theta +  \left[(\ell + a \cos\theta)^2 \sinh^2\alpha + \frac{\chi^2}{\sin^2\theta} \cosh^2\alpha \right] \delta^\mu_t \delta^\nu_t
+\frac{1}{\sin^2\theta} \,\delta^\mu_\phi \delta^\nu_\phi  \nonumber \\[2mm] &&
+ \left(\delta^\mu_t \delta^\nu_\phi + \delta^\mu_\phi \delta^\nu_t\right) \frac{\chi}{\sin^2\theta} \,\cosh\alpha - h^{\mu \nu}(\ell + a \cos\theta)^2.
\label{kilcommetr}
\end{eqnarray}
For the vanishing boost parameter $ \alpha =0 $, this expression  reduces to  the Killing tensor for the original Kerr-Taub-NUT spacetime (see, for instance, \cite{cglp}).

Next, we need to calculate  the nonvanishing components of the ``current vector"   in (\ref{cur}). Using equations (\ref{confmet}) and (\ref{conftens}), with equation (\ref{omega}) in mind,  it is not  difficult to show that the  current one-form is given by
\begin{eqnarray}
I & = &\frac{\sinh^2\alpha}{ B \Sigma^2} \left\{(\ell + a \cos\theta)^2\left[ M(2r^2-\Sigma) + 2 r \ell (\ell + a \cos\theta)\right] dr
\right. \nonumber \\[2mm]  & & \left.
+\, a  r^2 \sin\theta \left[(\Sigma-2 r^2) \ell + 2 M r (\ell + a \cos\theta)\right]d \theta \right\}\,.
\label{curvect}
\end{eqnarray}
Here we have also used the $ K_r^r $  and $ K_\theta^\theta $ components of the Killing tensor in (\ref{kilcommetr}).  It is interesting to note that  this expression does not  vanish  for $ a=0 $. In other words, even the Schwarzschild-Taub-NUT spacetime fails to  retain its Killing tensor  in the Kaluza-Klein framework. Thus, the hidden symmetries  of the original Kerr-Taub-NUT spacetime, generated by the irreducible Killing tensor, fail to survive from the Kaluza-Klein viewpoint.  Meanwhile,  the  hidden symmetries generated by the conformal Killing tensor do exist, guaranteeing  the complete separability of variables in the Hamilton-Jacobi equation for massless geodesics. With (\ref{omega}) in mind, it is straightforward to verify that the conformal Killing tensor in equation (\ref{conftens}), in which the Killing tensor is obtained by  lowering (with respect to the effective metric $  h_{\mu\nu} $) the contravariant indices of the tensor in (\ref{kilcommetr}), satisfies equation (\ref{confkil}) with the current vector given in   (\ref{curvect}).

\section{Conclusion}

The purpose of this Letter was to address the question of how and to what extent the hidden symmetries of the original Kerr-Taub-NUT spacetime survive in the Kaluza-Klein framework. Exploring  the separability structure of the Hamilton-Jacobi equation in the Kerr-Taub-NUT-Kaluza-Klein spacetime, we have found that the complete separation of variables  occurs only for massless geodesics. That is, the spacetime admits the hidden symmetries generated by the conformal Killing tensor. Using a simple trick built up  on  the conformally related metric with the Killing tensor,  we  have constructed the explicit expression  for the conformal Killing tensor. It should be noted that our separation trick, involving the conformally related metric with the Killing tensor, will also work for a wide class of rotating black hole solutions in four and higher dimensions, where the  complete separation of variables occurs for massless  fields. In particular, it will work for four-charge  rotating black holes of four-dimensional string theory \cite{cclp1} as well as  for three-charge  rotating black holes in five-dimensional ungauged $ \mathcal{N}=2 $ supergravity \cite{cclp2}. We hope  to turn to these issues in future works.

\section{Acknowledgment}

The author wishes to thank A. N. Aliev for  suggesting this problem and for very helpful discussions.  This work is supported by Istanbul Univesity Scientific Research Project (BAP) No. 9227.

\end{document}